\documentclass[prx,aps,twocolumn]{revtex4}
\usepackage{multirow}
\usepackage{amssymb}
\usepackage{amsmath}
\usepackage{mathrsfs}
\usepackage{graphicx}
\usepackage{amsfonts}
\usepackage{color}
\usepackage{bm}
\usepackage{xspace,soul}
\usepackage{hyperref}
\usepackage{graphicx,subfigure}

\newcommand{\bra}[1]{\ensuremath{\left\langle#1\right|}}

\newcommand{\ket}[1]{\ensuremath{\left|#1\right\rangle}}

\definecolor{darkgreen}{rgb}{0.0, 0.5, 0.0}

\begin{document}

\title{ Rise and fall of a bright soliton in an optical lattice }

\author{Piero Naldesi$^1$
\footnote{Electronic address: \texttt{piero.naldesi@lpmmc.cnrs.fr}}, 
Juan Polo Gomez$^1$, 
Boris Malomed$^{2,3}$, 
Maxim Olshanii$^{4}$, 
Anna Minguzzi$^1$ and 
Luigi Amico$^{5,6,7,8,9}$}

\affiliation{$^1$ Universit\'{e} Grenoble-Alpes, LPMMC, F-38000 Grenoble, France and CNRS, LPMMC, F-38000 Grenoble, France}
\affiliation{$^2$ Department of Physical Electronics, School of Electrical Engineering, Faculty of Engineering, Tel Aviv University, P.O.B. 39040, Ramat Aviv, Tel Aviv, Israel}
\affiliation{$^3$ Center for Light-Matter Interaction, Tel Aviv University, P.O.B. 39040, Ramat Aviv, Tel Aviv, Israel}
\affiliation{$^4$ Department of Physics, University of Massachusetts Boston, Boston, MA 02125, USA}
\affiliation{$^5$ Dipartimento di Fisica e Astronomia, Via S. Sofia 64, 95127 Catania, Italy}
\affiliation{$^6$ Centre for Quantum Technologies, National University of Singapore, 3 Science Drive 2, Singapore 117543, Singapore}
\affiliation{$^7$ MajuLab, CNRS-UNS-NUS-NTU International Joint Research Unit, UMI 3654, Singapore}
\affiliation{$^8$ CNR-MATIS-IMM \& INFN-Sezione di Catania, Via S. Sofia 64, 95127 Catania, Italy}
\affiliation{$^9$ LANEF \textit{'Chaire d'excellence'}, Universit\`e Grenoble-Alpes \& CNRS, F-38000 Grenoble, France}

\date{\today}

\begin{abstract}
We study an ultracold atomic gas with attractive interactions in a one-dimensional optical lattice. We find that its excitation spectrum displays a quantum soliton band, corresponding to $N$-particle bound states, and a continuum band of other, mostly extended, states. For a system of a finite size, the two branches are degenerate in energy for weak interactions, while a gap opens above a threshold value of the interaction strength. We find that the interplay between degenerate extended and bound states has important consequences for both static and dynamical properties of the system. In particular, the solitonic states turn out to be protected from spatial perturbations and random disorder. We discuss how such dynamics implies that our system effectively provides an example of a quantum many-body system that, with the variation of the bosonic lattice filling, crosses over from integrable non-ergodic to non-integrable ergodic dynamics, through non-integrable non-ergodic regimes.
\end{abstract}

\maketitle



\textit{ Introduction and summary of the results.}
 Solitary waves in classical fluids may arise when wave dispersion effects are compensated by non-linear interactions \cite{landau1959fluid}. The study of their mathematical properties has defined important domains in mathematical research \cite{Zakharov_1974,ablowitz1981solitons,faddeev2007hamiltonian}, with far reaching implications for pure and applied modern science \cite{belashov2006solitary,belinski2001gravitational,agrawal2000nonlinear,mei1989theory,scott1995solitary}. Although quantum mechanics is an intrinsically linear theory, solitons may emerge also in quantum fluids: in this case, non-linearities do arise as a result of effective cooperative phenomena occurring in quantum-many-particle systems. Indeed, solitons were demonstrated to emerge in different quantum mechanical contexts, ranging from quantum material science to particle physics \cite{lee1981particle,El06,leggett2006quantum,Horng09,Sacha14,Streltsov11}.

In this work, we are primarily motivated by the recent investigations in quantum fluids as provided by ultracold atoms trapped in one dimensional optical potentials. In these systems, bright solitons may emerge for attractive atom-atom interactions \cite{ablowitz1981solitons} (see also \cite{Carr_2000}) and they have been observed in several experiments \cite{strecker2002formation, khaykovich2002formation, Nguyen422, marchant2013controlled}. From a conceptual point of view, however, bosonic systems with attractive interactions must be treated with care: because of the Bose statistics, the lowest energy state can be macroscopically occupied with a density that is \textit{magnified} by interactions. Important progress has been achieved describing the bosonic fluid through a famous integrable theory as proposed by the Lieb-Liniger model, which is amenable to an exact analysis \cite{Lieb_1963}. Relying on that, it was demonstrated that the ground state energy may display instabilities that can be nonetheless cured by a suitable choice of interactions and density \cite{mcguire1964study}. Indeed, the limits of vanishing interaction with a finite density, leading to mean-field results for large number of bosons \cite{piroli2016local}, or of vanishing density with finite interactions \cite{calabrese2007correlation}, have been thoroughly explored. In particular, by analysing the solutions of the mean-field Gross-Pitaevskii equation it was found that \cite{Kanamoto_2003,kanamoto2005symmetry} a critical value of the attraction exists for which the ground state density undergoes to a transition from a uniform profile to a bright-soliton-type one (with the logic implied by the onset of modulational instabilities in the condensate). In the same limit, it was exactly demonstrated using the Bethe-Ansatz solution that density-density and higher-order correlation functions display a qualitative change of behaviour in correspondence to the critical value predicted by the mean-field theory. 

In this work we focus on a bosonic system described by the Bose-Hubbard Model (BHM) \cite{Fisher1989,Zol1998} \eqref{BHH} confined in a one dimensional lattice, where density and interactions strength are \textit{both} finite. We perform a numerical study of energy bands and quantum correlations using the Density Matrix Renormalization Group (DMRG) method \cite{whitedmrg,Daley04,Vidal03,white93, WhiteFeiguin2004}. Among different aspects implied by the lattice, here we exploit the energy-band structure of the system, featuring characteristic bendings, foldings and energy gaps. Such effects can indeed define new physical regimes in our system with peculiar bound states of solitonic type (see \cite{oelkers2007ground,Sorensen_2012,Barbiero_2014}).

For finite attractive interactions we find that $N$-body bound states are formed; for weak interactions, however, these bound states are degenerate with a second band of other states, mostly extended or involving lower-order bound states. By increasing the interaction strength, the bound states get more and more energetically favourable, until a critical interaction strength $U_c$ for which the band of bound states is completely separated by an energy gap from the rest of the spectrum.

\begin{figure}[h]
\centering
\includegraphics[width=1.0 \columnwidth]{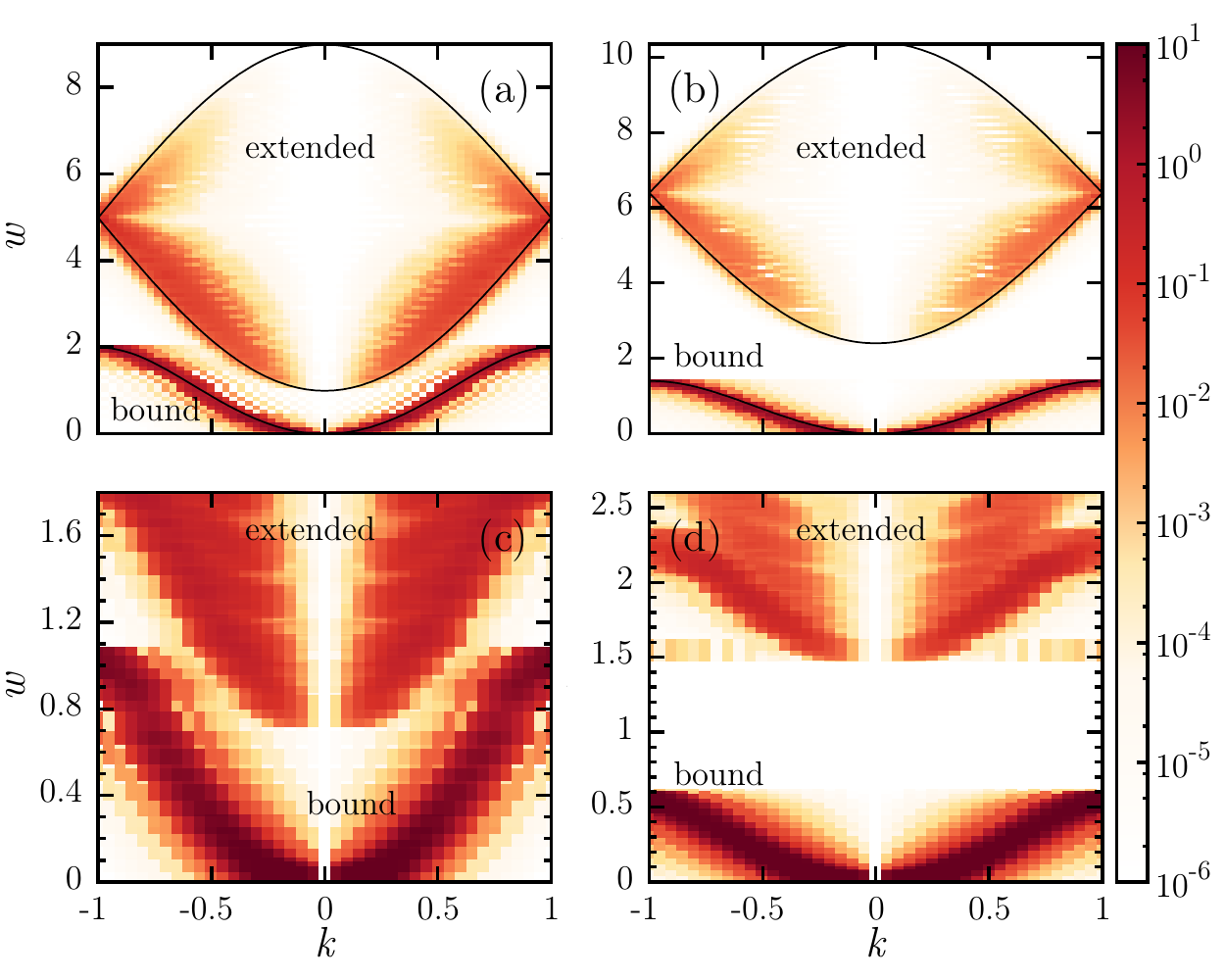}
\caption{Dynamical structure factor $S(k,\omega )$ for a chain of $L\!=\!30$ sites. Upper row: analytical results for $N\!=\!2$ particles for $U\!=\!2\!<\!U_{c}$ \textbf{(a)} and $U\!=\!5\!>\!U_{c}$ \textbf{(b)}. The black lines, obtained from the exact solution for $N=2$, outline two bands of bound (lower) and scattering (upper) states \cite{Valiente_2008, Boschi_2014}. Lower row: $S(k,\omega )$, numerical results for $N\!=\!5$ particles. In panels~\textbf{(c)} and ~\textbf{(d)} interaction are set to $U\!=\!0.75\!<\!U_{c}$ and $U\!=\!1.2\!>\!U_{c}$ respectively.}
\label{dynamical}
\end{figure}

\begin{figure}[h]
\centering
\includegraphics[width=1.0 \columnwidth]{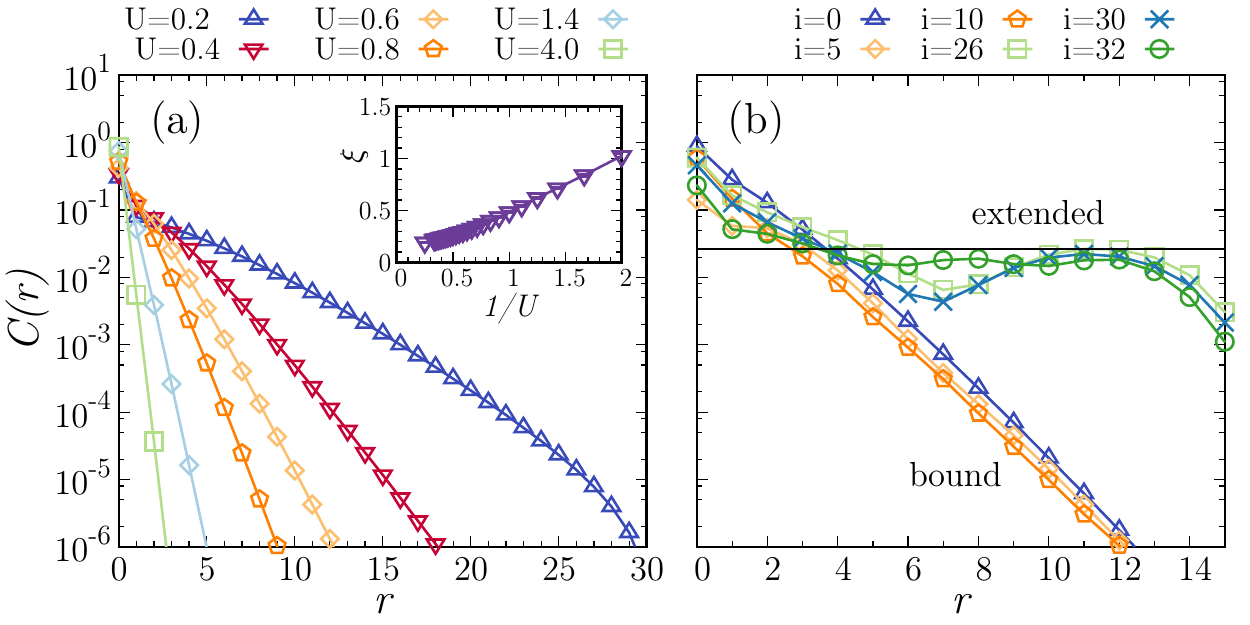}
\caption{ Density-Density correlation function $C(r)$ for $N\!=\!5$ particles in a chain of $L\!=\!61$ sites. Panel \textbf{(a)}: $C(r)$ for the ground state as a function of $U$. Inset: the correlation length $\xi$. Panel \textbf{(b)}: $C(r)$ calculated over several excited states for $U\!=\!0.6\!<\!U_{c}$, i labels the i-th excited state (i$=\!0$ correspond to the ground state) for a chain of $L=30$ sites.}
\label{correlations}
\end{figure}

The calculation of the dynamical structure factor \eqref{strfac} provides the portrait of the band structure of the system, displaying a two-branch dispersion at low energies as shown in Fig.~\ref{dynamical}. Such quantity is experimentally detectable by means in ultracold-atoms experiments by Bragg scattering methods \cite{fabbri2015dynamical,landig2015measuring}. We characterize the nature of ground and excited states in the spectrum by monitoring the density-density correlations functions, which display a different spatial behaviour for extended and $N$-particle bound states, see Fig.~\ref{correlations}.

\begin{figure}[h]
\centering
\includegraphics[width=1.0 \columnwidth]{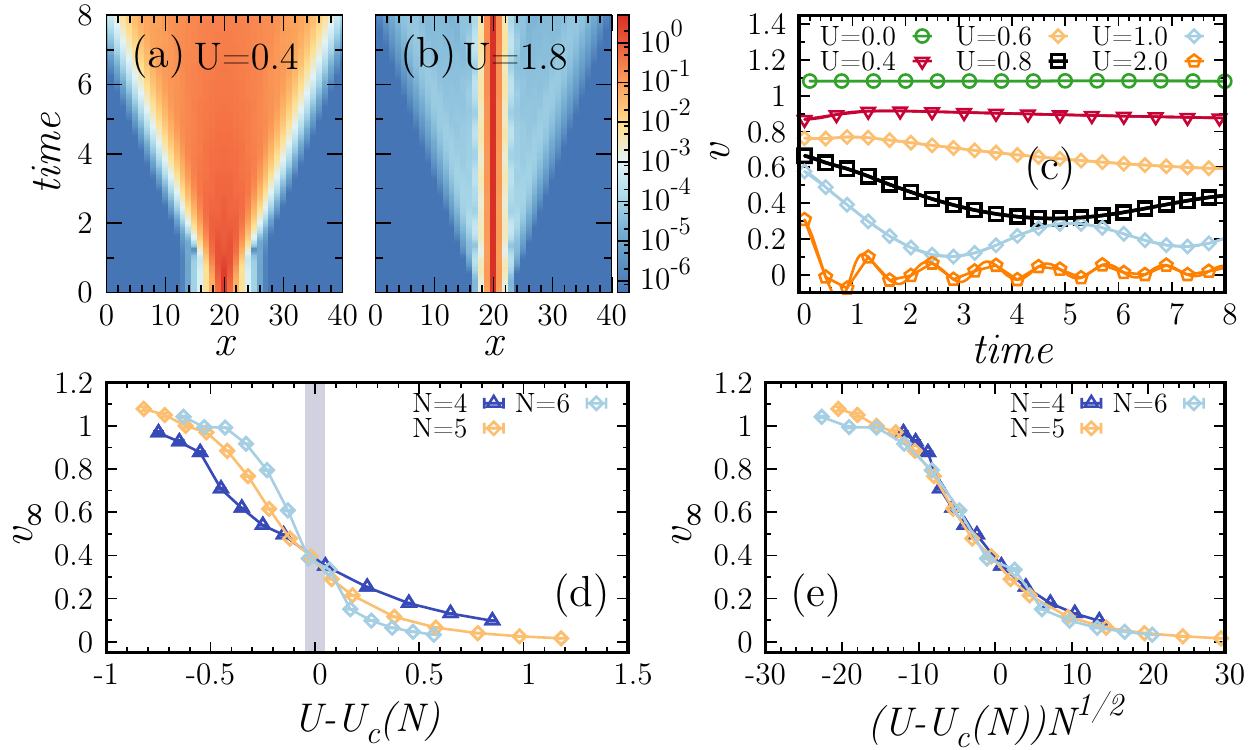}
\caption{Upper row \textbf{(a-b)} : expansion of a soliton composed by $N\!=\!5$ particles, pinned to the center of a chain with $L=41$ sites for different regimes. Panel \textbf{(c)}: expansion velocity $v(t)$ for different interaction strengths. The black line divides the gapless (upper) regions from the gapped (lower) ones. Panel \textbf{(d-e)}: asymptotic expansion velocity $v_{\infty }$ as a function of $U-U_c(N)$ and of $(U-U_c(N))\sqrt{N}$.}
\label{velocity}
\end{figure}

The changing of band structure and the opening of a gap varying the interaction, has important consequences for the dynamics of the system. We devise a protocol in which we prepare a quantum soliton in the middle of the chain and then we let it expand under the guidance of Hamiltonian \eqref{BHH}. For $U\!=\!0$ only extended states are available in the dynamics; for $0<U\!<\!U_{c}$ the extended and bound states are available; for $U\!>\!U_{c}$, at low energies, solely bound states exist. As a striking feature, in this latter regime the density keeps more and more the localized shape of the initial state and only a small fraction of the state spreads over the lattice. The crossover between the two regimes is purely mesoscopic since  $U_c$ scales like the inverse of the number of particles (see supplemental material). We note that the expansion velocity experiences a crossover from a large value for $U\!<\!U_{c}$ to a smaller value for for $U\!>\!U_{c}$. These features, that have been predicted for two particles \cite{Boschi_2014}, clearly emerge in Fig.~\ref{velocity}. 
Even though no quantum phase transition occurs in the system, we find that the expansion velocity close to $U_c$ displays  a universal scaling. Such a feature illustrates  the subtle interplay between interactions  and particle number in the dynamics of attractive bosons in a lattice. 

\begin{figure}[h]
\centering
\includegraphics[width=1.0 \columnwidth]{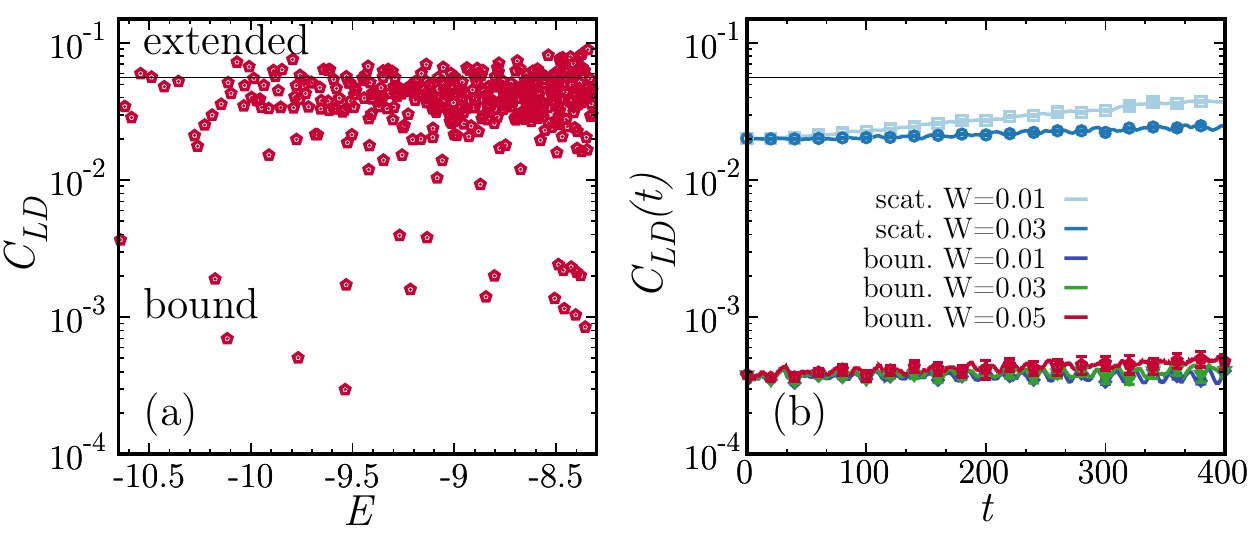}
\caption{
Panel \textbf{(a)}: expectation of $C_{LD}$ for several eigenstates $\phi _{n}$ as a function of their energy $E$, for $N\!=\!5$ particles in a chain of $L=21$ sites. Interaction is set to $U\!=\!0.5<U_{c}$. The black line at $C_{LD}=(N/L)^{2}\approx 0.056$ is a guide to the eye. Panel \textbf{(b)}: time evolution of $C_{LD}$, from scattering or bound state with adjacent energy eigenvalues.}
\label{NonErgodicCorrelations}
\end{figure}

\begin{figure}[h]
\centering\includegraphics[width=1.0\columnwidth]{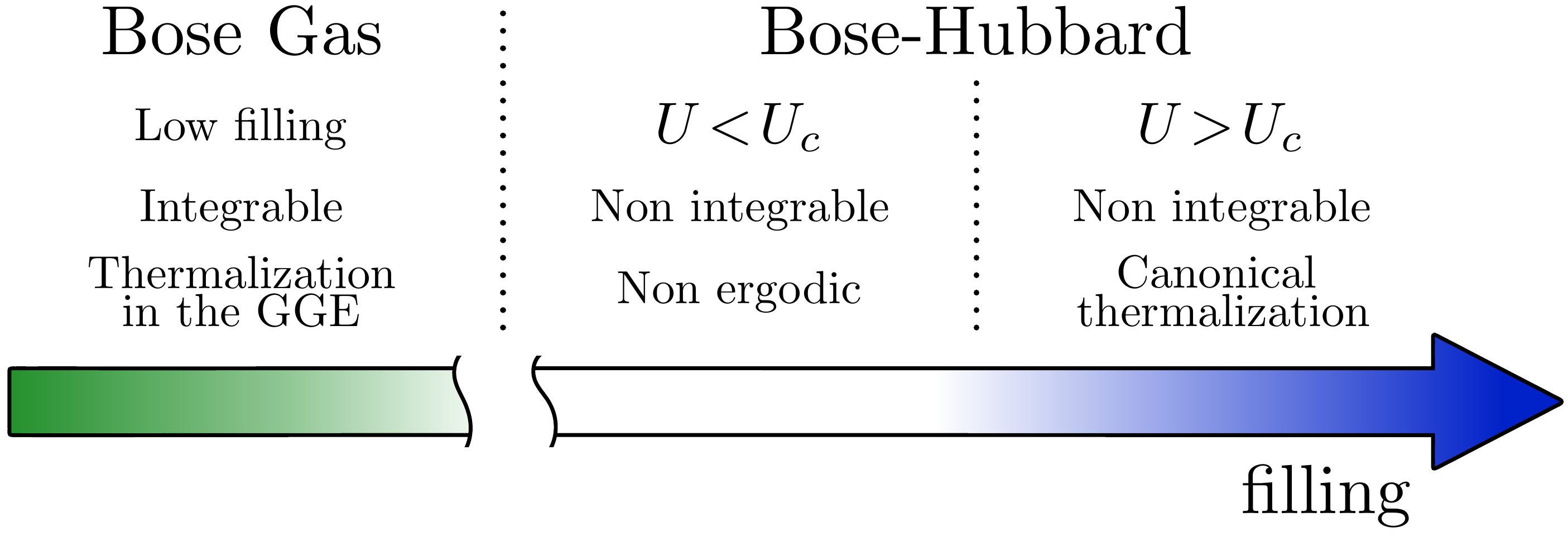}
\caption{Schematic diagram of the system as a function of the filling at varying
interaction strength. }
\label{qkam}
\end{figure}

We also find that the occurrence of degenerate scattering and bound states in the spectrum implies nontrivial time evolution of correlations: the large distance asymptotic density-density correlations are not function solely of the energy, but they strongly depend on the choice of the initial state. Indeed, not even a random perturbation is able to turn the time asymptotics of solitonic bound states into the one of scattering states, see Fig.~\ref{NonErgodicCorrelations}. Such a result indicates that bright solitons in the lattice are robust to external perturbations. This specific lack of ergodicity has implications at a fundamental level to study the interplay between thermalization and integrability, see Fig.~\ref{qkam}. 


\textit{ The model.}
\label{model} We employ the Bose-Hubbard Model (BHM) describing $N$ interacting bosons in a one-dimensional lattice. The Hamiltonian reads
\begin{equation}
\hat{\mathcal{H}}=-J\sum_{j=1}^{L}\left( a_{j}^{\dagger }a_{j+1}+\text{h.c.}%
\right) -\frac{|U|}{2}\sum_{j=1}^{L}\hat{n}_{j}\left( \hat{n}_{j}-1\right)
\label{BHH}
\end{equation}%
where the operators $n_{i}=a_{i}^{\dagger }a_{i}$ count the number of bosons at the site $i$; the operators $a_{i}$, $a_{i}^{\dagger }$ obey the canonical commutation relations $[a_{i},a_{j}^{\dagger }]=\delta _{ij}$, and $L$ is the number of sites. The parameters $J$, $U$ in \eqref{BHH} are the hopping amplitude and the strength of the on-site interaction, respectively. Throughout this article, we will consider only attractive interactions; both energies and the parameter $U$ are in units of $J$. Times are in units of $\hbar/J$. Equation \eqref{BHH} describes a closed system and therefore it neglects 3-body losses. We will also always (unless stated otherwise) consider open boundary conditions: $a_{1}^{\dagger }a_{L}=a_{L}^{\dagger }a_{1} = 0$. The BHM \eqref{BHH} is not solvable by the coordinate Bethe ansatz. The failure results because of finite probabilities that a given site is occupied by \textit{more than two} particles, whose interaction cannot be factorized in 2-body scattering \cite{haldane1980solidification, choy1980some, choy1982failure}.
Nevertheless the plane-wave ansatz of coordinate Bethe Ansatz works well in the so called
two-particle sector, for which such probabilities are vanishing\cite{Valiente_2008, Boschi_2014}.
Despite the fact that the BHM is not integrable, its continuous limit is the Bose-gas
integrable field theory \cite{amico2004universality} (see Supplemental Material). 
We note that a similar logic works also for discretization for the classical non-linear-Schroedinger equation \cite{kevrekidis2009discrete}. 


\textit{ Bound versus scattering states.}
\label{S}
Information on the available excitations in the system as a function of their momentum $k$ and energy $\omega$ is provided by the dynamical structure factor $S(k,\omega )$:
\begin{equation}
S(k,\omega )=\sum_{\alpha \neq 0}\sum_{r}|\left\langle \alpha \right\vert e^{-i k r} \hat{n}_{r}\left\vert 0\right\rangle |^{2}\delta (\omega -\omega _{\alpha }).
\label{strfac}
\end{equation}
where $\hat{n}_{r}$ is the number operator acting on the site $r$, $\ket{0}$ is the ground state and $\alpha$ labels the states with increasing energy (ie $\alpha=1$ is the first excited state).
The peaks of $S(k,\omega )$ reconstruct the energy bands of the system~ \cite{Mattis_1986,Valiente_2010}. We observe that for small $U$ a low-energy band separates from the rest of the spectrum. Corroborated by the exact results obtained for the BHM in the two-particle \cite{Valiente_2008, Boschi_2014} and three-particle \cite{Mattis_1986, Valiente_2010} sectors (see Supplemental Material), we conclude that for a general number of particles the lower band is always made of $L$ bound states. Such conclusion is further supported by the study of correlation functions presented below. Remarkably, because of the lattice bending of the energy bands, we observe that for $U\!<\!U_{c}$ the two bands of bound and extended states are partially overlapping (see Fig.~\ref{dynamical} (a,c)). For $U\!>\!U_{c}$ the two bands are fully separated by an energy gap which linearly grows with the interaction strength (see Fig.~\ref{dynamical}(b,d)). Further details on $U_c$ and a full study of the energy bands can be found in the Supplemental Material.
We point out that the regime of degeneracy between extended and bound states cannot be captured by neither mean-field nor continuous Lieb-Liniger model since both theories describe the solitonic states with a single parameter (the interaction $U$), without reference to any specific feature of the energy bands.

In order to characterize bound states, we study the density-density correlation function (being the system translationally invariant, the density itself does not display fruitful features): $C(r)\!=\!\left\langle n_{L/2}\:n_{L/2+r} \right\rangle$.
From our numerical analysis we confirm that the ground state is a bound state:
$C(r)\!\sim\!\exp (-r/\xi)$ 
with correlation length $\xi$ decreasing with increasing $U$ - Fig.~\ref{correlations} (a). For excited states, the lowest excitation branch is, indeed, made of bound states characterized by $C(r)$ decaying exponentially with a single $\xi $ depending solely on $U$. On the other hand, for states belonging to the second branch, at intermediate distances, $C(r)$ approaches a plateau $\sim n_{as}\!=\!(N/L)^{2}$, before dropping down when approaching the walls of the box. We thus can conclude that the higher branch contains extended states. This strong difference in the correlations in the two bands can be quantified by studying the density-density correlations at large distance, as defined by the correlator
$C^{LD}\!=\!\sum_{i=2}^{L/4}C(i)/\mathcal{N}$,
where $\mathcal{N}$ is the number of sites over which this function has its support. While for scattering states $C^{LD}\approx n_{as} $, see Fig.~\ref{correlations} (c), for bound states the magnitude of the correlations is several orders of magnitude smaller because of the faster decay of the corresponding $C(r)$. Such feature provides a clear indicator of the nature of the states (extended or bound).

We proceed to study the stability of these states through a suitable dynamical protocol.
Specifically, we address the evolution
of bound and scattering excited states, $\left\vert \psi
^{B}\right\rangle $ and $\left\vert \psi
^{S}\right\rangle $ respectively, with adjacent energy eigenvalues in the
spectrum for $U\!<\!U_{c}$, after
having perturbed the system by adding a random-noise source in it. 
The dynamics is then governed by a Hamiltonian
which depends both on the interaction strength $U$ and on the intensity of
the perturbation, $W$:
$\mathcal{H}(W,U)\!=\!\mathcal{H}(U)+\sum_{i}\epsilon _{i}n_{i}$, where $\epsilon _{i}$ is a random variable chosen uniformly in the interval $[-W,+W]$. In Fig.~\ref{NonErgodicCorrelations} (b) we note that (within the time scale available in our numerical simulations) $C^{LD}(t)$ remains almost constant in the course of the evolution, meaning that bound and scattering states are not mixed by random disorder. This is a strong evidence, on the timescale considered, of the soliton stability and robustness.

\textit{ Dynamical expansion of pinned solitons.}
Finally, we devise a specific dynamical protocol to evidence the features of the band structure shown in  Fig.~\ref{dynamical}. 

A soliton is pinned in a given site $i_{0}$ of the lattice, by initially breaking the lattice translational symmetry with an attractive potential $\mathcal{H}_{i}(\mu ,U)\!=\!\mathcal{H}(U)+\mu (U)n_{i_{0}}, \label{pinn-main}$ and then let it expand. The pinning energy $\mu(U)$ is chosen such that the energy injected in the system by the perturbation is equal to the width of the bound-state
band (see Supplemental material). In this way, while for small $U$ we populate both scattering and bound states, for $U\!>\!U_{c}$ when the gap separates the two bands, mostly bound states are populated.

The dynamical evolution is governed by $\mathcal{H}(U)$ obtained by removing the pinning potential. 
In Fig.~\ref{velocity} (a-b-c) we show the expansion dynamics of the density
for three cases: $U\!<\!U_{c}$, $U\!\approx\!U_{c}$ and $U\!>\!U_{c}$. At increasing the
interaction strength, we see that the density profile stays
closer and closer to the shape of the initial state, only its small
fraction spreading into the chain. This can be seen more quantitatively by
studying the expansion velocity: $v(t)\!=\!(d/dt)\sqrt{R^{2}(t)\!-\!R^{2}(0)}$, with $R^{2}(t)\!=\! (1/N)\sum_{i=1}^{L}n_{i}(t)\left( i\!-\!i_{0}\right)^2$. 
In Fig.~\ref{velocity} (d-e) we show, respectively, $v(t)$ and its asymptotic value, $v_{\infty }$, at large times. 

While for $U<U_c$ no oscillations are visible within the simulation time considered, at increasing $U\geq U_c$ the velocity displays typical oscillations with period scaling as $\hbar/U$.

The asymptotic expansion velocity $v_{\infty }$ is identified by fitting it to a phenomenological expression, $v(t) \!\approx\! v_{\infty}+\cos (At)/t^{B}$, where $A$ and $B$ are fitting parameters. The inspection of $v_{\infty }$ in Fig.~\ref{velocity} (e) further shows the difference between the two regimes. 

Interestingly enough, close to $U_c$, we find that $v_{\infty }$ displays scaling behaviour; the results  are  not affected by the size of the system.  While there is no criticality in the system,  the observed feature is due to a diverging time scale  associated to the soliton thermalization: as critical slowing down implies scaling, here, the scaling is due to the fact that the soliton cannot equilibrate to the state with uniform density.

\textit{ Conclusions.}
\label{conclusions}
In this work, we studied the spatial correlations and dynamical properties of attractive bosons in one-dimensional lattices. The presence of the lattice induces a characteristic energy band structure, for which bright solitons display specific properties with distinctive correlation functions. Such features can affect the dynamics of the system substantially. We have demonstrated how a bright solitonic bound state can be created in the system and, by studying the expansion dynamics, we have provided a way to test its stability against external perturbations. 

Our work can be relevant for 
fundamental studies on the ergodicity of quantum systems. 
Thermalization in quantum many-body systems is usually expressed
in terms of the well-known Eigenstates Thermalization Hypotesis (ETH) \cite%
{Srednicki1994, Deutsch1991}: if the expectation values of local observables
for individual eigenstates are a smooth function of energy, then
the system behaves ergodically and one can exchange, for such observable,
the long-time average by the Gibbs ensemble average with no memory of the specific initial state except its energy. 
The bimodal distribution of correlations of Fig~\ref{NonErgodicCorrelations} (a), (b) with co-existence, for $U\!<\!U_{c}$, of the two {families} of states at the same energy leads to a clear violation of the ETH. Our understanding of the system can be summarized in Fig~\ref{qkam}, where three regimes may occurr.

{\small I)} Since \textit{lattice spacing $\Delta $ results to be vanishing 
proportionally to the filling factor $\nu\!=\!N/L$}, at sufficiently small $\nu$, our bosonic system is integrable
(described by the Bose-gas field theory\cite{korepin1997quantum,amico2004universality}). According to the general
theory, in this limit the system is expected to thermalize to a Generalized Gibbs
Ensemble. \newline
{\small II)} By increasing the filling facctor, the system does not remain 
integrable, being described by the Bose-Hubbard model. For such system, when
bound states and scattering states coexist with equal energies
(i.e. for $U\!<\!U_{c}$), the long-times asymptotic states strongly depend on
the initial states.
\newline
{\small III)} At larger filling, the system is far from integrability,
as the Bose-Hubbard corrections to the Bose-gas grow stronger. In this case, the solitonic band is nearly flat, making the coexistence between bound and scattering states impossible. In this limit, therefore, the system is ergodic.
Such a scenario indicates \textit{that going from {\small I)} to
{\small III)}, the integrability, controlled by the filling (instead
of the perturbation added to the Hamiltonian, in the
framework of more standard approaches), is destroyed by entering an
intermediate regime, in which the system keeps some trace of integrability in that the dynamics is not ergodic.} In this sense then, we contribute to the search of a quantum analog of the KAM theorem \cite{rasetti} which is one of the key challenges in contemporary research (see f.i. \cite{reichl1987search,chang1987quantum,Caux2015}). We note that the  scenario emerging in our work   is in line with the findings of Rigol~\cite{rigol2009breakdown}. As a follow up of the present study, it would be interesting to study  whether more complex bound states and observables at  higher energies can provide different types of intermediate thermalizations. 

We believe that our analysis is within the current activity in atomic physics quantum technology. Ultracold atoms with tunable interactions have been already loaded in one-dimensional optical lattices in several experiments \cite{lewenstein2007ultracold,Heinze13,Greif15,Meinert16}. A specific protocol allowing to address bound and scattering states selectively at a given mean energy can be implemented by quenching the interaction from repulsive or attractive interactions and evolving with the target Hamiltonian \cite{Naldesi2017}. Atomtronic circuits can also provide appropriate tools to explore the dynamics of the system \cite{seaman2007atomtronics, amico2005quantum, Amico_Atomtronics, Amico_NJP}.

\begin{acknowledgments}
\textit{Acknowledgments}. We want to thank E. Ercolessi, F. Ortolani and L. Taddia for helpful discussions. The Grenoble LANEF framework (ANR-10-LABX-51-01) is acknowledged for its support with mutualized infrastructure. We thank National Research Foundation Singapore and the Ministry of Education Singapore Academic Research Fund Tier 2 (Grant No. MOE2015-T2-1-101) for support. We thank ANR SuperRing (ANR-15-CE30-0012-02). B.M. and M.O. acknowledge support provided by an integrated program in physics of NSF and Binational (US-Israel) Science Foundation through grants No. PHY-1402249 and PHY-1607221 (NSF), and 2015616 (BSF). 
\end{acknowledgments}


\begin{thebibliography}{10}
\bibitem{footnote_edge_states}
For open boundary conditions two edge states appear at the border of the chain.
  For $U>U_c$ these two states have energy between the two bands of scattering
  and bound state \cite{Pinto_09}.
\end{thebibliography}

\clearpage
\newpage
\appendix

\begin{center}
{\bf SUPPLEMENTARY MATERIAL for} \\
{\bf \emph{Rise and fall of a bright soliton in an optical lattice.}}
\end{center}

\section{Derivation of the continuum model}
We sketch the derivation of the Bose Gas and Lieb- 
Liniger models as continuous limit of the Bose-Hubbard quantum dynamics.

Let's define the density of bosons in the lattice as $D=N/(L \Delta)$,
$\Delta$ being the lattice spacing. In the continuous limit $\Delta \rightarrow 0$, which implies that the filling factor $\nu=N/L=D \Delta$ must be accordingly small.
In the continuous limit the bosonic operators must be rescaled: $a_i=\sqrt \Delta \Psi (x)$,
 $n_i= \Delta \Psi^\dagger (x)\Psi (x) $, $x=\Delta i$.
Then, the Bose-Hubbard model reduces to the Bose gas quantum field theory \cite{korepin1997quantum}:
$H_{BH}=t\Delta^2 {\cal H}_{BG}$,
$
{\cal H}_{BG}=\int dx \left [(\partial_x \Psi^\dagger)(\partial_x \Psi)
+ c \Psi^\dagger\Psi^\dagger\Psi\Psi \right ]
$, 
with $c=U/(t\Delta)$\cite{amico2004universality}.
The quantum fields obey $[\Psi (x),\Psi^\dagger (y)]=\delta(x-y)$ and
$[\Psi^\dagger (x),\Psi^\dagger (y)]=0$.
The Bose gas field theory is the quantum field theory for the Lieb-Liniger model. Such statement can be demonstrated by writing the eigenstates of
${\cal H}_{BG}$ as $|\psi({\mathbf{\lambda}})=\int d {\mathbf z} \chi({\mathbf z}|{\mathbf{ \lambda}}) \Psi^\dagger (z_1)\dots \Psi^\dagger (z_N)|0\rangle$.
Then, it can be proved that $\chi({\mathbf z}|{\mathbf{ \lambda}})$ must be eigenfunctions of
\begin{equation}
H_{LL}=-\sum_{j=1}^L \frac{\partial^2}{\partial z_j^2}-2c\sum_{L\ge j> k \ge 1} \delta (z_j-z_k) \,.
\end{equation}
%
%
In the mean field limit, the Lieb-Liniger model reduces to the Gross-Pitaevskii equation (see f.i.~ \cite{Calogero_1975})
\begin{equation}
\left[-{{\partial^2}\over{\partial x^2}} -2cN\left ( 1-{{1}\over{N}}\right ) |\phi(x)|^2\right ] \phi(x)=\mu \phi(x) \;.
\label{GPE}
\end{equation}
The Gross-Pitaevskii equation above results to be the stationary solution of the
functional $E[\phi] =\int d {\mathbf z} \chi_{GS}^*({\mathbf z}|{\mathbf{ \lambda}}) H_{LL} \chi_{GS}({\mathbf z}|{\mathbf{ \lambda}}) $
in which the ground state ansatz of $H_{LL}$: $\chi_{GS}({\mathbf z}|{\mathbf{ \lambda}})=(1/N)\prod_j^N \phi(x_j)$ was adopted ($\mu$ is the normalizing constant for $ \phi(x)$).

We note that the integrability of the continuous theory can indeed be preserved through regularization, but the resulting lattice Hamiltonians do contain higher order non-linearities \cite{amico2004universality,dutta2015non}.

\section{Two-particle exact solution of the attractive BHM}
We discuss some of the existing results for the few particle Bose-Hubbard model that we have used as benchmarks for our studies of the $N>2$ particle sector of the BHM. Since the Hamiltonian (1) commutes with the total number of particles, we can study the model separately in every sector with well defined number of particles.

The two-particle problem is perfeclty integrable since only 2-particle interaction are possible. In this sector one can separate the center-of-mass and the relative coordinates and use the standard Bethe ansatz technique to provide an exact solution for the Schr\"odinger equation.
As it was found by \cite{Valiente_2008,Boschi_2014}, for a finite attractive interaction the spectrum decomposes in two bands: one band of scattering states and a second one composed by bound states which are characterized, respectively, by real and complex relative momenta. The band of bound states is composed by exactly $L$ states, one for each center-of-mass momenta.
It has also been pointed out that there exists a critical value of the interaction $U_c=4$ for which both bands are completely separated in energy.
Note that for any $U\neq0$ the ground state is always a bound state however, for $U<U_c$ some of the excited bound states occur to be degenerate in energy with some states in the scattering band. Only for $U>U_c$ the two bands are fully separated by an energy gap that grows linearly with the interaction.

In Fig.~\ref{spectrum} (a),(b) of the main text we plot $S(k,\omega)$ in the 2-particle case for $U=3$ and $U=5$, which are respectively below and above the critical value $U_c$. In this image we also include the envelope of the scattering band and the energy of the bound state band found with the Bethe-Ansatz solution \cite{Boschi_2014}.

The three-particle problem is already too complex to be completely solved but still the full energy spectrum can be obtained \cite{Mattis_1986,Valiente_2010}. Whereas the characterisation of all the states of the system is really complex, a band of 3-particle bound states at low energy can still be separated from the rest of the spectrum when interactions are larger than a critical value, that is different from the one of 2-particle case.

\section{Numerical studies of the spectral properties of the attractive BHM}
\label{spectrumANDcorrelations}
\begin{figure}[tbp]
\centering
\includegraphics[width=1\columnwidth]{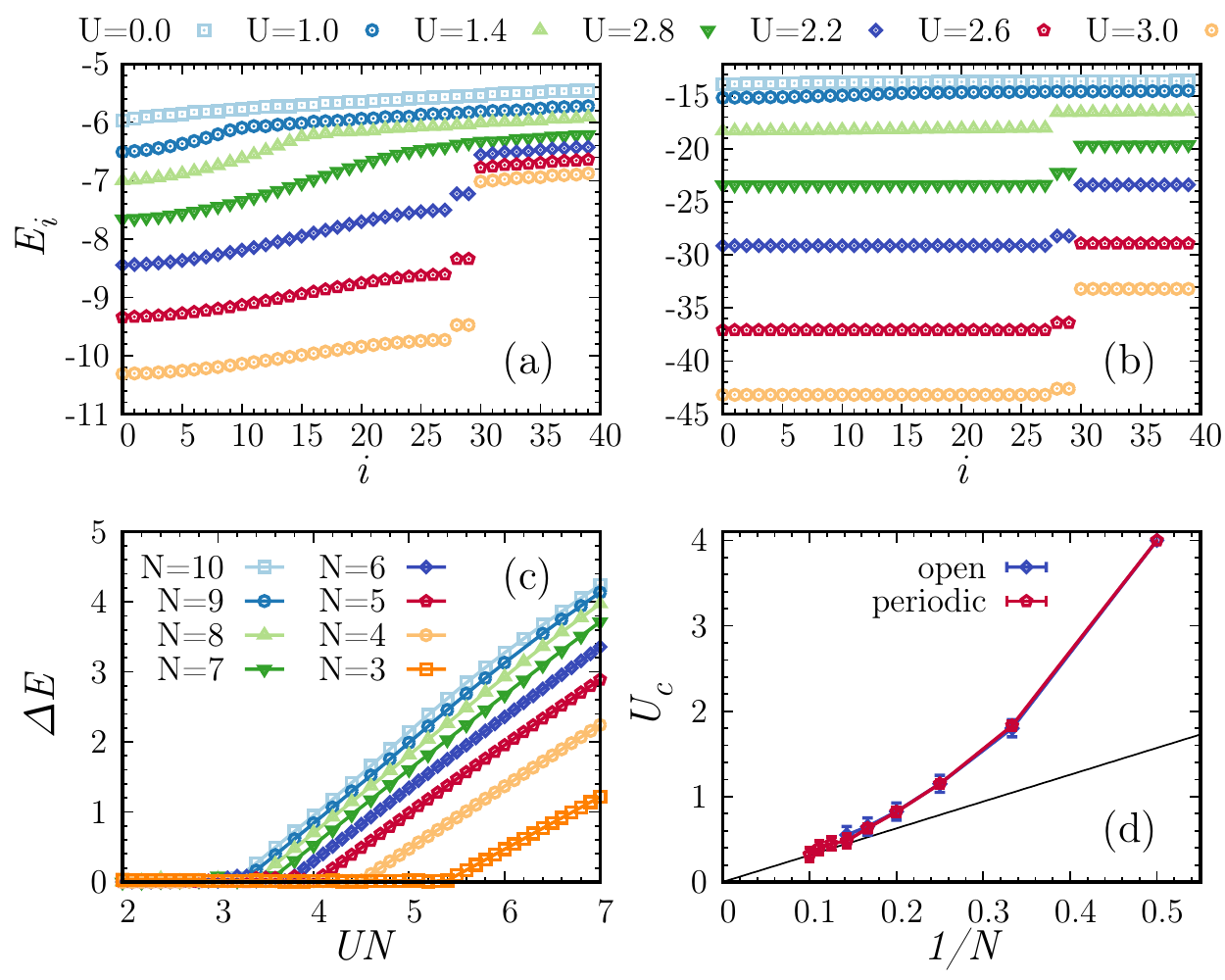}
\caption{Energy spectrum of the attractive Bose-Hubbard model,
defined as per Eq. (1). Upper row: energies
of the first 40 eigenstates for $N=3$ \textbf{(a)} and $N=7$ \textbf{(b)}
particles in the box (i.e. with open boundary conditions) of $L=30$ sites
calculated for different values of the interaction strength
$U$, as produced by the DMRG analysis of the excited states
of the system. While for small $U$ the energy is a continuous function of
the state index $i$, for $U>U_{c}$ and energy gap separates the first $L$
states from the rest of the spectrum. In this regime, we also note two edge
modes, with an energy intermediate between the two bands, triggered by the
presence of open boundary conditions in the chain. Lower row:
\textbf{(c)} energy gap $\Delta E$ as a function of $UN$ for different $N$,
calculated in the chain of $L=30$ sites with open boundary conditions.~ \textbf{(d)} Critical interaction strength $U_{c}$ needed to open an energy gap between the
band of bound state and the rest of the spectrum, as a
function of $1/N$ in the lattice with $L=30$ sites. The black line $1/N$ is
a guide to the eye. The transition between the two regimes
is not senstive to the geometry of the system, taking place
for the same $U_{c}$ both for open and periodic boundary conditions.}
\label{spectrum}
\end{figure}

This appendix is dedicated to the energy spectrum of the attractive BHM for the $N>2$ particle sector.
In Fig.~\ref{spectrum} (a),(b) of this appendix we show that a critical value of $U$ exists for which a
finite bandgap opens in the spectrum. Specifically, we show the energy of
the first 40 exited states of the lattice of $L=30$ sites with open boundary
conditions (hard-wall box), for $N=3$ and $N=7$ particles.

The energy gap between the bands is usually defined as the difference in energy between the last bound state, i.e. the $L$-th excited state, and the next one: $\Delta E=E_{L+1}-E_{L}$. For open boundary conditions, special care has to be taken due to the presence of the two edge states appearing at the border of the chain \cite{footnote_edge_states} (doublets in Figs.~\ref{spectrum}(a) and (b)) of the appendix -- we therefore adopt an alternative definition, $\Delta E=E_{L+1}-E_{L-2}$. The inspection of the gap as a function of a rescaled interaction strength, $NU$, in Fig.~\ref{spectrum} (c) of the appendix, allows us to conclude that there exists a critical value of the interaction strength, $U_{c}$, for which a emerging bandgap is displayed in the spectrum. In particular, the behaviour of $U_{c}$ in Fig.~\ref{spectrum} (d) of the appendix shows that the gap-opening mechanism is general and does not depend on the choice of boundary conditions.

\section{Critical pinning and soliton band width}
\label{appcrit}
\begin{figure}[!!!htt!!!]
 \centering
 \includegraphics[width=1 \columnwidth]{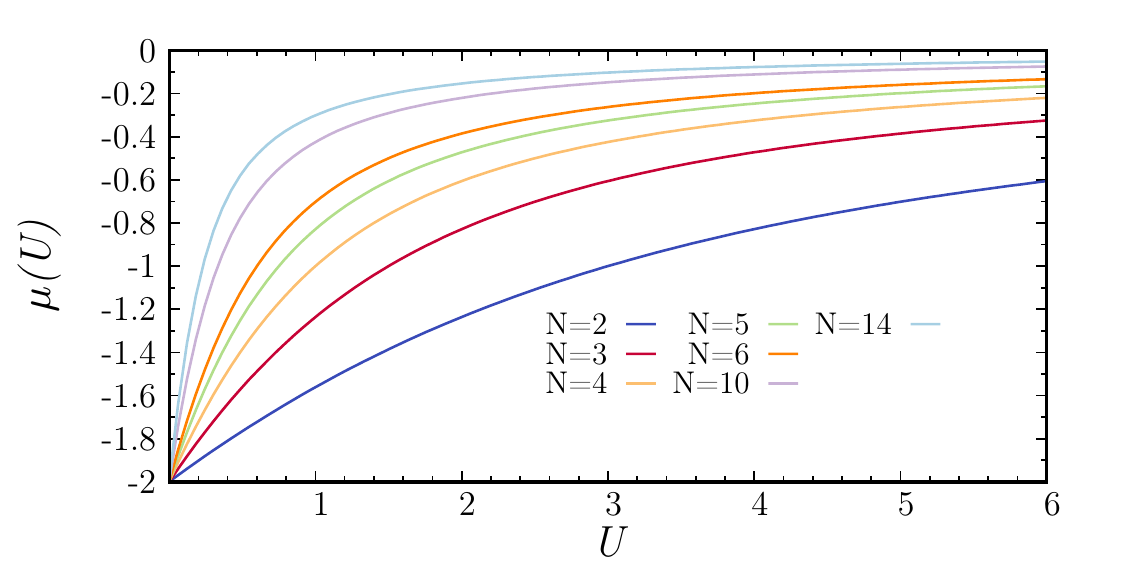}
 \caption{$\mu(U)$ from Eq. \ref{pin} for different number of particles $N$ and interaction $U$}
 \label{pino}
 \end{figure}

In this appendix we describe how to pin the initial state (in space) while keeping a large projection over the bound states of the system. The pinning procedure, which breaks the translational symmetry of the system by adding an attractive local potential, is modeled by the total Hamiltonian
\begin{equation}\label{pinn}
{\cal H}_i(\mu,U) = {\cal H}(U) + \mu(U) n_{i_0}
\end{equation}
whose ground state $\psi^i_0$ will be the initial state of our dynamics and where ${\cal H}(U)$ is given by the BH Hamiltonian Eq.~(1). Since the soliton we want to prepare is fully localized in space and bound states are, at fixed momentum, the states with lowest energy, the minimum requirement is to populate all of them.
We will therefore require that the energy we are introducing in the system (described by ${\cal H}(U)$) with the pinning must be equal to the bound state band width $\Delta E_B(U)$:
\begin{eqnarray}
\epsilon &=& \langle \psi^i_0 | {\cal H}(U) | \psi^i_0 \rangle - \langle \psi_0 | {\cal H}(U) | \psi_0 \rangle \\
&=& \mu(U) N = \Delta E_B(U)
\label{pinning_critical}
\end{eqnarray}
where $\psi_0$ is the ground state of ${\cal H}(U) $. $\Delta E_B(U)$ can be estimated from a general expression \cite{Valiente_2010} as:
\begin{eqnarray}
\Delta E_B &=&
\sqrt{\mathcal{U}^2+\mathcal{K}_{min}^2} -
\sqrt{\mathcal{U}^2+\mathcal{K}_{max}^2}\\
&=& \mathcal{U} - \sqrt{\mathcal{U}^2+\mathcal{K}_{max}^2}
\label{potkin}
\end{eqnarray}
where $\mathcal{U}$ is the potential energy and $\mathcal{K}_{max}$, $\mathcal{K}_{min}$ are respectively the maximum and the minum of the kinetic energy over the band. The second step in \ref{potkin} follows from $\mathcal{K}_{min}=0$.
The contribution to the potential energy can be estimated as:
$
\mathcal{U}=-\frac{|U|}{2}N(N-1)$ and $\mathcal{K} = 2N
$. 
With these assumptions, Eq. (\ref{pinning_critical}) yields
\begin{equation}
\mu(U) = \frac{|U|(1-N)}{2}\left( 1 - \sqrt{1+\frac{16}{U^2(N-1)^2}}\right)
\label{pin}
\end{equation}
In Fig.~\ref{pino} of this appendix we plot $\mu(U)$ for different number of particles $N$ and interaction $U$.

We observe that for a large value of $N$, $\mu(U)$ tends to zero. This is compatible with the idea that, for large $N$, the kinetic energy of the solitonic state is negligible compared to the potential one. In this case, the band becomes {effectively} flat and all the states are degenerate.

By introducing a large attractive pinning, all the particle will populate only one site. This state will have an energy $UN(N-1)/2$, which corresponds to the upper limit of the soliton band and it has been proven for $N = 2$ particles to have \cite{Boschi_2014} the largest projection over the band of bound state.\\

\section{Density matrix renormalization group}
Here we provide specific details of the numerical simulations, performed using Density matrix renormalization group (DMRG) techniques, presented in our paper for the energy, correlation functions, dynamical structure factor both for open and periodic boundary conditions as well as of the time-dependent simulations.

The DMRG is an numerical technique that hallows to study both statics and dynamical properties of general quantum systems in one dimension \cite{whitedmrg, white93, WhiteFeiguin2004}.

Since in our paper we focused on system of particles with attractive interactions we never cut the local Hilbert space in our simulations, allowing all the $N$ particles to occupy every single site of our lattice. This requirement makes the Hilbert grow really fast as $\approx N^L$, where $L$ is the number of sites. Moreover increasing the number of particles $N$ or the strength of the interaction massive degeneracies occurs in the spectrum (see Appendix \ref{appcrit}). These two difficulties make the simulations extremely challenging and clearly reduces up to $N \approx 5 \sim 10$ particles in systems with $L \approx 50 \sim 100$ lattice sites.

The study of the excited states id performed by building the density matrix of the system as an equal-weight mixture of the first l+1 states $\ket{\phi_i}$, where l is the number of highest excited state we want to calculate.
\begin{eqnarray}
\rho_l = \frac{1}{l+1} \sum_{i=0}^l \ket{\phi_i}\bra{\phi_i}
\label{potkin}
\end{eqnarray}
Through the DMRG procedure, $\rho_l$ is then project over a reduced Hilbert space with dimension $\mathcal{D}_{\mathcal{H}}$.  
In our work we calculate up to $l=400$ excited state. In order to limit the truncation error below $10^{-8}$ at each step we keep $\mathcal{D}_{\mathcal{H}} = 2000$.\\
For the time-dependent simulations of the non-ergodicity and expansion, the time-evolution is based on a Runge-Kutta 4th order scheme, with a time step of $\Delta t = 0.01$. For those cases we kept a variable number of states (up to 1500 for the simulations at larger times) allowing us to keep
the truncation error smaller than $10^{-6}$ at each time step. All time-dependent simulations have been performed imposing open boundary conditions in the lattice.

\end{document}